\documentclass[singlecolumn,showpacs,%
aps,superscriptaddress,
prd,notitlepage,showkeys,
nofootinbib,floatfix]{revtex4-1}

\usepackage{ulem}
\usepackage{amssymb}
\usepackage{amsmath}
\usepackage{graphicx}
\usepackage{dcolumn}
\usepackage[colorlinks,urlcolor=blue,citecolor=blue,linkcolor=blue]{hyperref}
\usepackage{color,units}
\usepackage[dvipsnames]{xcolor} 
\usepackage{lineno}
\usepackage{xspace}
\usepackage{longtable} 
\usepackage{float} 
\usepackage{multirow}
\usepackage{mathtools}
\usepackage{amsfonts,wasysym,epsfig,verbatim,subfigure,bm,mathrsfs,lipsum}
\begin{document}
\newcommand{\OKC}{The Oskar Klein Centre, Department of Astronomy, Stockholm University, AlbaNova, SE-10691 Stockholm,
Sweden}
\newcommand{\UHH}{Universit\"{a}t Hamburg, D-22761 Hamburg, Germany}
\newcommand{\AUT}{Aristotle University of Thessaloniki, Department of Physics
, 54124 Thessaloniki, Greece}

\newcommand{\bhaskar}[1]{{\color{black} #1}}

\title{A Bayesian investigation of the neutron star equation-of-state vs. gravity degeneracy} 
\author{Bhaskar Biswas}\email{phybhaskar95@gmail.com}
\affiliation{\OKC}\affiliation{\UHH}
\author{Evangelos Smyrniotis}
\affiliation{\AUT}
\author{Ioannis Liodis}
\affiliation{\AUT}
\author{Nikolaos Stergioulas}
\affiliation{\AUT}
\date{\today}

\begin{abstract}
Despite its elegance, the theory of General Relativity is subject to experimental, observational, and theoretical scrutiny to arrive at tighter constraints or an alternative, more preferred theory. 
In alternative gravity theories, the macroscopic properties of neutron stars, such as mass, radius, tidal deformability, etc. are modified. This creates a degeneracy between the uncertainties in the equation of state (EoS) and gravity since assuming a different EoS can be mimicked by changing to a different theory of gravity. We formulate a hierarchical Bayesian framework to simultaneously infer the EoS and gravity parameters by combining multiple astrophysical observations. We test this framework for a particular 4D Horndeski scalar-tensor theory originating from higher-dimensional Einstein-Gauss-Bonnet gravity and a set of 20 realistic EoS and place improved constraints on the coupling constant of the theory with current observations. Assuming a large number of observations with upgraded or third-generation detectors, we find that the $A+$ upgrade could place interesting bounds on the coupling constant of the theory, whereas with the LIGO Voyager upgrade or the third-generation detectors (Einstein Telescope and Cosmic Explorer), the degeneracy between EoS and gravity could be resolved with high confidence, even for small deviations from GR.
\end{abstract}

\maketitle

\section{Introduction}
With the first detection of a gravitational wave (GW) signal from a binary neutron star (BNS) merger event GW170817~\cite{TheLIGOScientific:2017qsa} using the advanced LIGO \cite{LIGO} and advanced Virgo \cite{Virgo} detectors, simultaneously with its electromagnetic counterpart \cite{LIGOScientific:2017ync}, we have now entered the era of multimessenger astronomy (MMA) of neutron stars (NSs) with GWs \cite{Branchesi2018}. Novel constraints on the NS equation of state (EoS) \cite{LATTIMER2016127,RevModPhys.89.015007,Baym_2018} can be obtained from the inspiral phase of a binary's signal as it carries the imprint of NS matter due to the tidally deformed structure~\cite{Hinderer:2007mb,Binnington:2009bb,Damour:2009vw,Chatziioannou2020,Dietrich2021} of the components. Apart from the GW observations, recently the NICER collaboration~\cite{2016SPIE.9905E..1HG} has also provided constraints on masses and radii of PSR J0030+0451~\cite{Miller:2019cac,Riley:2019yda} and PSR J0740+6620~\cite{Miller:2021qha,Riley:2021pdl}, by observing X-ray emission from hot spots on the NS surface. By combining these observations from multiple messengers it is possible to arrive at stringent constraints on NS properties~\cite{Raaijmakers:2019dks,Jiang:2019rcw,Most:2018hfd,Traversi:2020aaa,Xie:2019sqb,Landry_2020PhRvD.101l3007L,Biswas:2020puz,Biswas:2021yge,Al-Mamun:2020vzu,Dietrich:2020efo,Miller:2021qha,Raaijmakers:2021uju,Biswas:2021paf}. However, all these studies have been performed under the assumption that general relativity (GR) is the true theory of gravity. However, there is a degeneracy between the NS EoS and the assumed model of gravity when the equilibrium structure of NSs is obtained, and the current uncertainties due to these two different effects largely overlap. Since the spacetime of NSs exhibits a strong curvature, one should examine the uncertainties in the EoS of NSs simultaneously within GR and other alternative theories of gravity. For current constraints on deviations from GR using GW observations, see \cite{LIGOScientific:2021sio}. Scheduled upgrades and new generation of GW detectors  are expected to yield a large number of BNS
detections, resulting in a high cumulative signal-to-noise ratio and improved constraints,  by combining the information from multiple events \cite{Abbott2016Livrev,PhysRevD.91.043002,PhysRevD.97.084014,PhysRevD.100.103009,Chatziioannou:2021tdi,2023arXiv230812378I,Mondal:2023gbf}.

\begin{figure}[ht]
    \centering
    \includegraphics[width=0.45\textwidth]{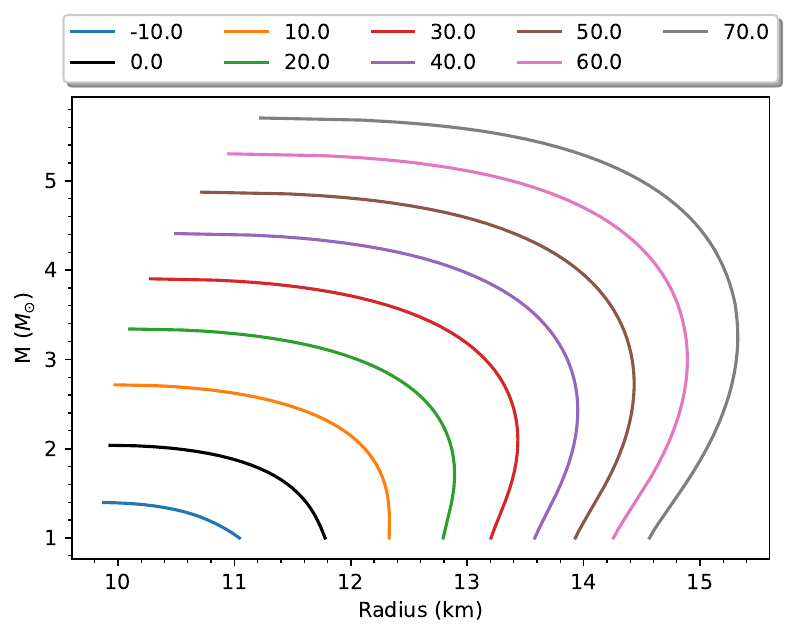}
    \caption{Representative mass vs. radius sequences for different values of the coupling constant $\alpha\,[\mathrm{km^2}]$ using the SLY4 EoS. }
    \label{fig:mr_sly_EGB}
\end{figure}

 In this paper, we consider a particular 4D Horndeski scalar-tensor model originating from higher-dimensional Einstein-Gauss-Bonnet (EGB) gravity. This theory contains several attractive properties/symmetries (see \cite{Charmousis:2021npl}), and its action takes the form
\begin{equation}
        S=\dfrac{1}{2\kappa}\int\mathrm{d}^{4}x\sqrt{-g}\left\{R+\alpha\left[\phi\mathcal{G}+4G_{\mu\nu}\nabla^\mu\phi\nabla^\nu\phi-4(\nabla\phi)^2\Box\phi+2(\nabla\phi)^4\right]\right\}+S_\mathrm{m},
\label{eq:action}
\end{equation}
where $\kappa=8\pi G/c^4$, $\mathcal{G}=R^2-4R_{\mu\nu}R^{\mu\nu}+R_{\mu\nu\rho\sigma}R^{\mu\nu\rho\sigma}$ is the Gauss-Bonnet scalar, and $S_\mathrm{m}$ is the matter Lagrangian. We take the scalar to be dimensionless; hence $\alpha$ has dimensions of length squared. In Ref.~\cite{Charmousis:2021npl}, the equilibrium of static, spherically symmetric, nonrotating NSs was studied in this theory of gravity (satisfying a set of modified Tolman–Oppenheimer–Volkoff (TOV) equilibrium equations) by modeling their interiors using tabulated EoS under the perfect fluid assumption. The mass-radius relations obtained by solving the modified TOV equations are shown in Figure~\ref{fig:mr_sly_EGB} for the SLY4~\cite{Douchin:2001sv} EoS for specific values of the coupling constant $\alpha$ in the range of $(-10,70) \,[\mathrm{km^2}]$ (with $\alpha = 0$ corresponding to the mass-radius sequence obtained in GR). Negative (positive) values of $\alpha$ lead to NS with larger (smaller) masses and radii, which is similar to varying the EoS within GR (hence the degeneracy).  

In this paper, we focus on investigating this  EoS-gravity degeneracy in the light of multimessenger or future GW observations of NSs. We formulate a hierarchical Bayesian framework in Section~\ref{section:methods and results} to simultaneously infer the EoS and population of NS and the coupling constant $\alpha$ by combining multiple astrophysical observations. To speed up the inference framework, we use a neural-network-based surrogate developed in an accompanying publication~\cite{Liodis}. First, we apply our inference framework using the current observational constraints and obtain posterior distributions for $\alpha$, under different prior assumptions mass population distributions of NSs. Next, we show that future GW detectors with improved sensitivity can potentially provide very tight constraints on the coupling constant $\alpha$, which could be sufficient to constrain the EoS and distinguish gravity from \bhaskar{GR simultaneously.} 
\bhaskar{At present, we are restricted to astrophysical constraints on tidal deformability and radii derived within the assumption of the validity of GR
Hence, we would like to emphasize that the present work focuses on the methodology of the hierarchical Bayesian framework, and quantitative results should be considered as indicative. We discuss the current limitations in detail in Section~\ref{section: discussion} and indicate how these can be improved in the future.}

\section{EoS Catalogue}

\begin{figure}[ht]
    \centering
    \includegraphics[width=0.45\textwidth]{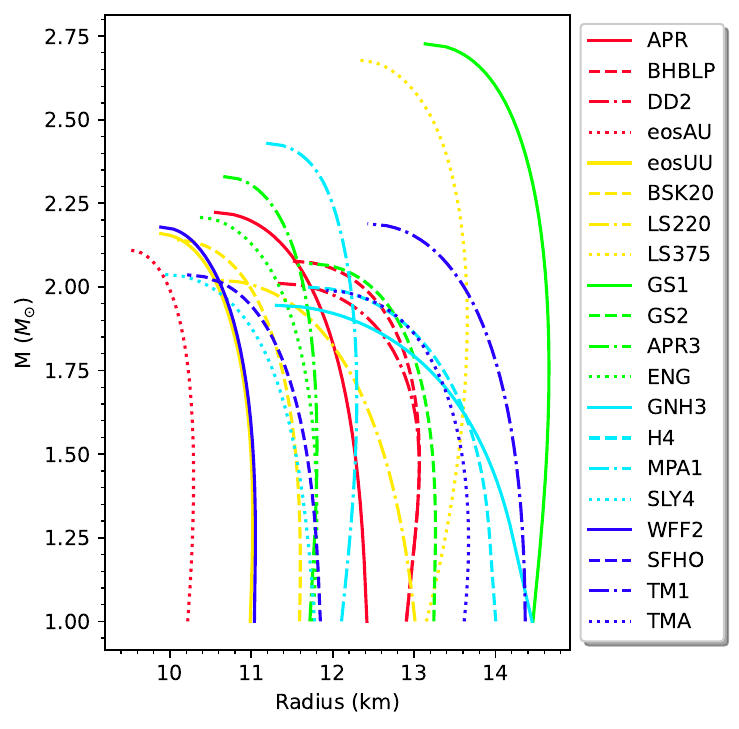}
    \caption{Mass vs. radius for sequences of nonrotating equilibrium models in GR for all the 20 EoSs considered in this study. 
    }
    \label{fig:mr_GR}
\end{figure}

We consider 20 EoS models, which are computed from different nuclear-physics approximations. This EoS collection has previously been used in Ref.~\cite{Bauswein:2020xlt} and comprises 
\bhaskar{(i) Variational-method EoSs (APR and APR3)~\citep{APR4_EFP}, (eosAU, eosUU, and WFF2)~\citep{wff1}, (ii) nonrelativistic mean field models with Skyrme interactions such as BSK20~\citep{Goriely_2010,Pearson_2011}, SLY4~\citep{Douchin:2001sv} and LS type model (LS220,LS375)~\cite{2012ascl.soft02011L}   which is very similar to a Skyrme apart that they assume a momentum independent interaction potential, (iii) relativistic Brueckner-Hartree-Fock EOSs MPA1~\citep{MPA1}, ENG~\citep{Engvik_1994}) (iv) relativistic mean field theory EoSs DD2~\cite{Hempel:2009mc}, (GS1, GS2)~\cite{Shen:2011kr}, SFHO~\cite{Steiner:2012rk}, TM1~\cite{Hempel:2011mk}, TMA~\cite{Hempel:2011mk}.} Also, we consider models with hyperons H4~\citep{H4} and BHBLP~\cite{Banik:2014qja}. In Figure~\ref{fig:mr_GR}, we plot the mass-radius sequences for each EoS assuming GR. The collection spans a wide region range of radii [10-14.3] km. \bhaskar{We intentionally include EoSs that are currently not favored in GR~\cite{Biswas:2021pvm}, but that can be made compatible within the 4D EGB gravity by choosing a suitable value of $\alpha$ (a positive value to increase the radius or a negative value to decrease the radius)\footnote{See also \cite{Biswas:2021yge} for the impact of experimental constraints on the determination of the EoS.}.}

\section{Artificial neural network surrogate model} This work focuses on inferring the NS EoS and 4D EGB gravity parameter $\alpha$ simultaneously by combining multiple astrophysical observations using Bayesian statistics. Performing this type of inference using Markov Chain Monte Carlo (MCMC)/nested sampling algorithms is very expensive. The main computational cost comes from solving the modified TOV equations presented in Ref.~\cite{Charmousis:2021npl} to calculate the mass and radius of an equilibrium model for a given EoS, as this process needs to be repeated over a million times to ensure well-converged posterior distributions of model parameters. The numerical solution of the modified TOV equations is based on an iterative method, which is far more expensive computationally than the direct Runge-Kutta-type integration used in GR. For this reason, in the accompanying publication~\cite{Liodis}, two types of Artificial Neural Network \bhaskar{(ANN)} surrogate models are built for each EoS: $f_1(\rm EoS, \alpha, p_c) \longrightarrow (M,R)$ and $f_2(\rm EoS, \alpha, M) \longrightarrow R$, where $p_c$ is the central pressure of a NS. The function $f_1$ takes EoS, $\alpha$, $p_c$ as input and returns the mass and radius of the corresponding NS,  whereas $f_2$ takes EoS and $M$ as inputs and returns the radius of the NS.  Both of these functions are highly accurate and offer speed-ups between $10-100$ times, depending on the input parameters. We use these functions in the inference computations in the next sections.

\section{Methodology and Results}
\label{section:methods and results}
\subsection{Constraints using current observations} 
\label{subsec:current constraints}
At first, we combine currently available measurements of NS's macroscopic properties made by astrophysical observations. We include mass-radius measurements of PSR J0030+0451~\citep{Riley:2019yda,Miller:2019cac}~\footnote{In the final stage of our manuscript preparation, an updated mass-radius measurement~\cite{Vinciguerra:2023qxq} of PSR J0030+0451 was published, which is consistent with previous constraints. This may have a small effect on the posterior distributions of $\alpha$, but we do not expect a major change in our results.} and PSR J0740+6620~\citep{Riley:2021pdl,Miller:2021qha} and the EOS insensitive posterior samples for the masses and radii of the two components of GW170817~\citep{TheLIGOScientific:2017qsa} and GW190425~\citep{LIGOScientific:2020aai}. 

From Bayes’ theorem, we can write the posterior probability distribution of $\theta$ for a set of astrophysical data $\{d\}$ as follows:

\begin{equation}
    p(\theta \mid \{d\}) = \frac{p(\{d\} \mid \theta)p(\theta)}{p(\{d\})} \,,
\end{equation}
where, $p(\{d\} \mid \theta)$ denotes the joint likelihood, which is defined as the multiplication of individual event likelihoods

\begin{equation}
    p(\{d\} \mid \theta)= \prod_{i} p(d_{i} \mid \theta) \,,
\end{equation}
and $p(\{d\})$ is the Bayesian evidence.
The individual likelihoods coming from  observations containing an isolated NS or from a binary system, which provide mass and radius measurements, take the following forms, respectively
\begin{widetext}
\begin{subequations} \label{eq:likelihood}
\begin{align}
    p(d|\theta) = \int^{m_{\mathrm{max} }}_{m_{\mathrm{min} }} dm \  p(m|\theta) 
    p(d| m, R (m, \theta)), \\
    p(d|\theta) = \int^{m_{\mathrm{max} }}_{m_{\mathrm{min} }} dm_1 dm_2 p(m_1,m_2|\theta) 
    p(d| m_1, R_1 (m_1, \theta), m_2, R_2 (m_2, \theta)),
\end{align}
\end{subequations}
\end{widetext}
where  $\theta$ includes a  parameter characterizing the EoS, the coupling constant $\alpha$ of the modified gravity theory, and additional mass model parameters. We assume that all NSs originate from a common mass distribution  $p(m \mid \theta)$ and form BNS with random paring: 
\begin{equation}
    p (m_{1}, m_{2} \mid \theta) \propto p(m_{1} \mid \theta) p(m_{2} \mid \theta) \Theta(m_{1} > m_{2}).
\end{equation}
We consider three models of NS mass distribution~\cite{Landry:2021hvl}:

\begin{widetext}
\begin{subequations} \label{ns_mass_models}
\begin{align}
    p_\textsc{u}(m|m_{\rm min} = 1,m_{\rm max}) :=& U(m|m_{\rm min},m_{\rm max}) , \label{flat} \\
    p_\textsc{n}(m|\mu,\sigma,m_{\rm min},m_{\rm max}) :=& \mathcal{N}(m|\mu,\sigma)U(m|m_{\rm min},m_{\rm max})/A , \label{peak} \\
    p_\textsc{nn}(m|\mu,\sigma,\mu',\sigma',w,m_{\rm min},m_{\rm max}) :=& \left[w \mathcal{N}(m|\mu,\sigma)/B + (1-w) \mathcal{N}(m|\mu',\sigma')/C\right] U(m|m_{\rm min},m_{\rm max}) , \label{bimodal}
\end{align}
\end{subequations}
\end{widetext}
subject to the constraint $m_{\rm min} \leq m \leq m_{\rm max}$ where both $m_{\rm min}$ and  $m_{\rm max}$ depend on  the EoS parameter and $\alpha$. \bhaskar{In Eq.~\ref{ns_mass_models}, $U$ and $\mathcal{N}$ stand for the uniform and Gaussian distributions, respectively.} For all the population models we fix $m_{\rm min}=1$.
In Eq.~\eqref{flat}, the normalization factor $1/(m_{\rm max}-m_{\rm min})$ acts as an Occam's razor, preferring the EoS with slightly larger $m_\mathrm{max}$ than the heaviest observed NS mass and disfavoring EoS with much larger $m_\mathrm{max}$.
The mass model in Eq.~\eqref{peak} assumes NS masses are distributed according to the Gaussian distribution, characterized by mean $\mu$ and standard deviation $\sigma$, which are fixed at $1.33 M_{\odot}$ and $0.09 M_{\odot}$, respectively. \bhaskar{The parameter values of the Gaussian distribution are taken from Ref.~\cite{Farrow:2019xnc}, where they were inferred by combining a total of 17 galactic double neutron star systems.}
The last mass model (Eq.~\ref{bimodal}) is the sum of two Gaussian distributions with fixed parameter values ($\mu=1.34$, $\sigma=0.07$, $\mu'=1.80$, $\sigma'=0.21$) and a fixed mixing fraction parameter $w=0.65$ that determines the relative weight of the Gaussian components. \bhaskar{The parameter values of the ``two Gaussians" distribution are taken from Table 3 of Ref.~\cite{Alsing:2017bbc}, where they were inferred by combining a total of 74 galactic NS mass measurements.} The normalization constants $A$, $B$, and $C$ in Eqs.~\eqref{peak}-\eqref{bimodal} ensure that $\int p(m|\theta) \, dm = 1$.

\begin{figure}[ht]
    \centering
    \includegraphics[width=0.45\textwidth]{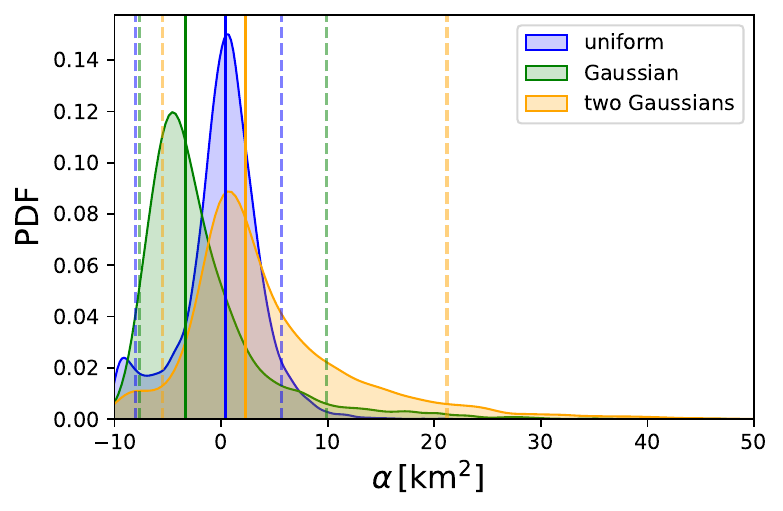}
    \caption{Marginalized posterior distribution of $\alpha$ obtained using current astrophysical observations under the assumption of three different NS mass population models. For each model, the corresponding median and $90 \%$ CI are also indicated using solid and dashed lines in the same color, respectively.}
    \label{fig:alpha_post_astro}
\end{figure}

\begin{table}[ht!]
\caption{Median and $90 \%$ CI of $\alpha$, log-evidences ($\log_{10} Z$) (point estimate and $1 \sigma$ CI) are quoted here for three different choices of the NS mass population model.   
}
\begin{tabular}{|c|c|c|c|} 

\hline
 Quantity&uniform & Gaussian & two Gaussians  \\
\hline 
$ 90 \% \, \mathrm{CI \, of} \alpha\,[\mathrm{km^2}]$
&$0.40_{-8.43}^{+5.29}  $  &$-3.36_{-4.27}^{+13.22}$  
&$2.30_{-7.82}^{+18.90}$ \\

\hline
$\log_{10} Z$
& $-10.82 \pm 0.02 $  
& $-31.90 \pm 0.02$ 
& $-9.25 \pm 0.04$ \\
\hline

\end{tabular}
\label{tab:current-result}
\end{table}

In this work, we focus on inferring the EoS parameters and the coupling constant $\alpha$, as we fix the parameters involved in the population models. Since we do not use a parametrized EoS but specific proposed EoS,  we assign an integer index number for each EoS candidate from the catalog. In our Bayesian framework, we consider a uniform prior on the EoS index, where a random draw from the prior will pick an integer number between 1 and 20. For $\alpha$, we choose a broad uniform prior of $(-10,70) \, \rm{[km^2]}$. We also need priors on the mass parameters of each observation, which are all uniform between $m_{\rm min}$ and $m_{\rm max}$. The posterior of these parameters is computed using a nested sampling algorithm implemented in~{\tt Pymultinest}~\citep{Buchner:2014nha}.  

In Figure~\ref{fig:alpha_post_astro}, the marginalized posterior distribution of $\alpha$ is shown after combining GWs and NICER observations of NSs under the assumption of three different choices of the mass population model. The corresponding median and $90 \%$ confidence interval (CI) of $\alpha$, log-evidences $\log_{10} Z$ are quoted in Table~\ref{tab:current-result}. In a previous study~\cite{Charmousis:2021npl}, assuming only a minimum black hole mass of $M = 5.7 \pm 1.8 M_{\odot}$ for the lighter component of GW200115, yields a conservative estimate, $\alpha = 285^{+207}_{-171} \, \rm{[km^2]}$ at $90 \%$ CI. If the lightest component in GW190814 with mass $2.59_{-0.09}^{+0.08} M_{\odot}$ was a black hole, then the constraint becomes $\alpha \lesssim 59 \mathrm{~km}^2$, and current constraints on masses and radii of neutron stars (obtained in a GR framework, however) yield an estimated constrain of $\alpha \lesssim 40 \mathrm{~km}^2$. In comparison to \cite{Charmousis:2021npl}, we obtain tighter constraints\footnote{See also \cite{Doneva:2020ped,Gammon:2023uss} and references therein for additional constraints on 4D EGB theories.} of $\alpha \lesssim 20 \mathrm{~km}^2$ (90\% CI) when using the \bhaskar{``two Gaussians"} population model\footnote{Notice that negative values of $\alpha$ can be excluded on the grounds that that atomic nuclei should not be shielded by a horizon, see \cite{Charmousis:2021npl}.}.

\bhaskar{The current astrophysical observations do not have enough power~\cite{Landry:2021hvl} to clearly favor one of the functional forms of NS mass distributions considered in this study. Nevertheless, for the particular distributions we consider (with parameters chosen as described above), we find some interesting results. Specifically, whereas the posteriors of $\alpha$ are consistent with each other within 90 \% CI, the log-evidence value is much lower for this particular Gaussian NS mass population model when compared to the uniform and ``two Gaussians" models}. Following the interpretation of Kass and Raftery~\cite{Bayes-factor}, we can decisively rule out a model if $\log_{10}\mathcal{B}_j^i \leq -2$, where \bhaskar{$\mathcal{B}_j^i := Z_i/Z_j$ is the {\it{Bayes factor}} between two models labeled $i$ and $j$}. Hence, \bhaskar{this particular} Gaussian population model is ruled out. This is not surprising, as the mass estimate of PSR J0740+6620 is $2.08 \pm 0.07 \, M_{\odot}$, which is outside of the $1 \sigma$ interval of the Gaussian mass population model. There is no statistically significant difference between the uniform and the ``two Gaussians" models. However, the ``two Gaussians" model is slightly preferred over the uniform model. The ``two Gaussians" model prefers a slightly higher value of $\alpha$ compared to the uniform model, as it has a slightly higher preference for higher mass due to the presence of a secondary peak (at $1.8 \, M_{\odot}$). As higher values of $\alpha$ lead to more massive NSs, this result is justified. We note that in light of current multimessenger observations, all models are consistent with GR.       

\begin{figure}[ht]
    \centering
    \begin{tabular}{cc}
    \includegraphics[width=0.45\textwidth]{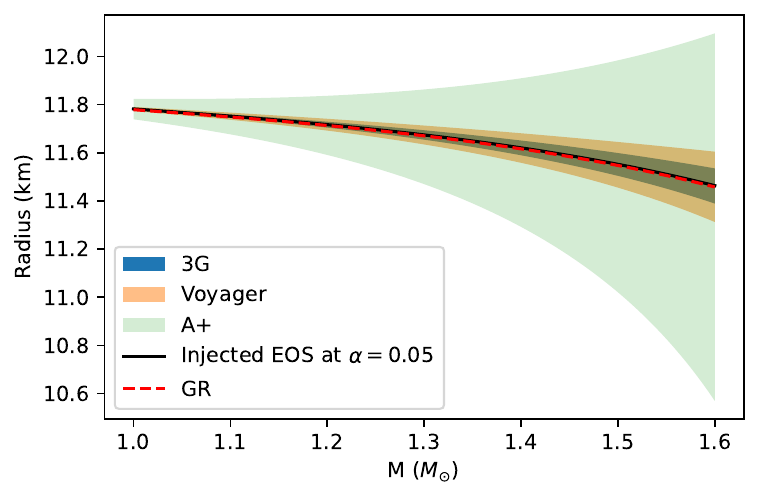} &
    \includegraphics[width=0.4\textwidth]{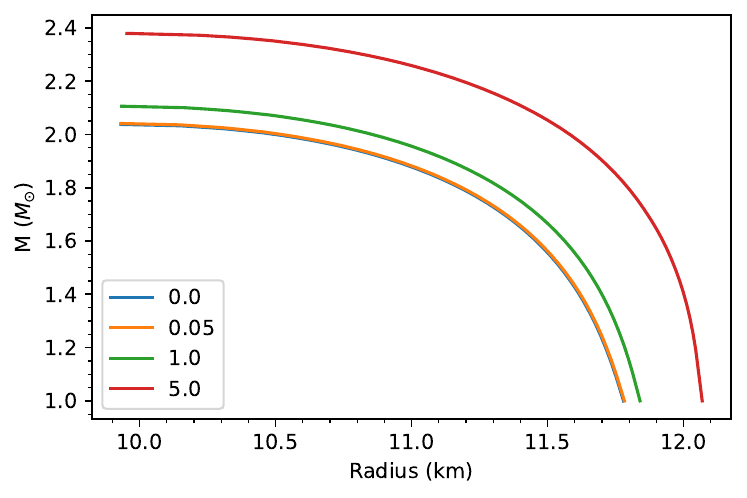} \\
    \end{tabular}
    
    \caption{{\it 
    Left panel}:  The uncertainty in radius as a function of mass in the [1.0,\,1.6] $M_\odot$ range at three different detector sensitivity levels. In this example, the injected EOS is SLY4 with $\alpha=0.05$. {\it 
    Right panel}: Mass vs. radius sequences for the different injected values of $\alpha$ (the GR case corresponds to $\alpha=0$).}
    \label{fig:uncertainy_estimation}
\end{figure}

\subsection{Constraints with future GW detectors} 
\label{seq:future}

Future GW detectors with improved sensitivity are expected to detect louder signals and thus will provide better overall constraints after multiple events of different masses are combined.  A single GW170817-like event would have an SNR ($\Tilde{\Lambda}$ $90\%$ uncertainty) of 100(200) at design aLIGO sensitivity \cite{aLIGOnoise}, 200(100) with A+~\cite{aLIGO_design_updated}, 300(66) with Voyager~\cite{LIGO:2020xsf}, and 1000(20) with 3G detectors (Einstein Telescope \cite{Maggiore:2019uih} and Cosmic Explorer \cite{Reitze:2019iox}). For this analysis, we adopt the estimates of 
 Chatziioannou~\cite{Chatziioannou:2021tdi} and
 assume a {\it total accumulated} SNR 200,  1000, and 10000 with A+, Voyager, and 3G dectectors, respectively (the latter estimate is based on \cite{Haster:2020sdh}).  Using similar considerations as in~\cite{Chatziioannou:2021tdi}, we plot the corresponding accuracy of NS radius as a function of mass in Figure~\ref{fig:uncertainy_estimation}. We use \bhaskar{SLY4} EoS and  $\alpha =[0.05,1,5] $ for the injection.
The smallest of these values,  $\alpha = 0.05$, is very close to GR. This is on purpose, as we want to test if future GW detectors will be able to detect even such small departures from GR. 

\begin{figure*}[ht]
\begin{center}
\begin{tabular}{ccc}
\includegraphics[width=0.3\textwidth]{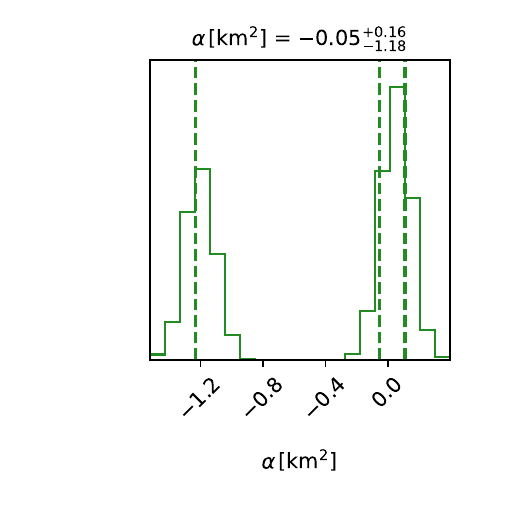}&
\includegraphics[width=0.3\textwidth]{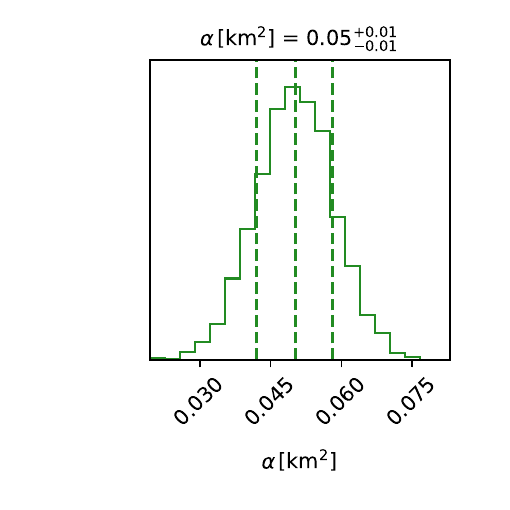}
\includegraphics[width=0.3\textwidth]{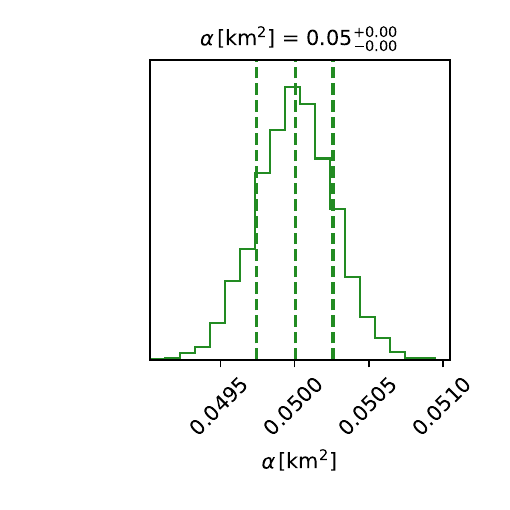} \\

\includegraphics[width=0.3\textwidth]{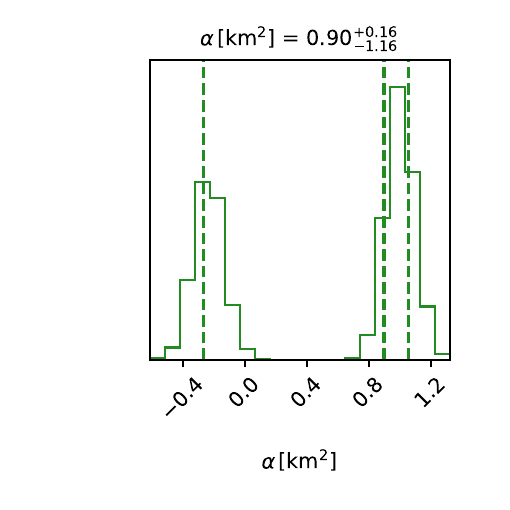}&
\includegraphics[width=0.3\textwidth]{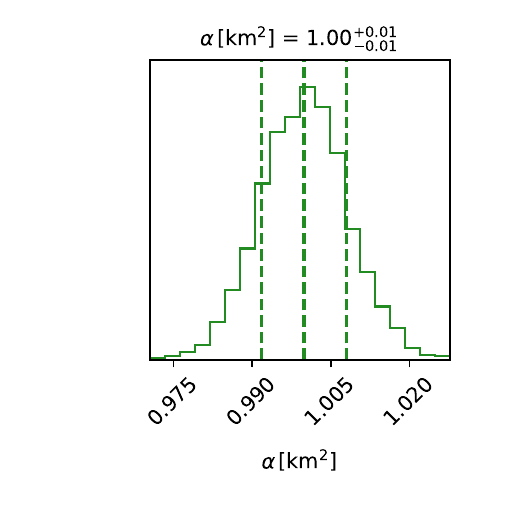}
\includegraphics[width=0.3\textwidth]{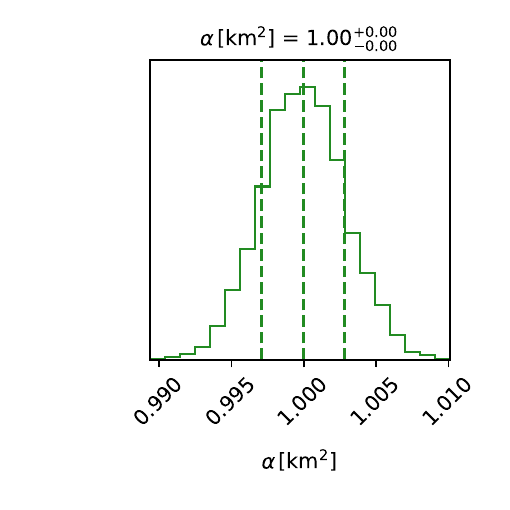} \\

\includegraphics[width=0.3\textwidth]{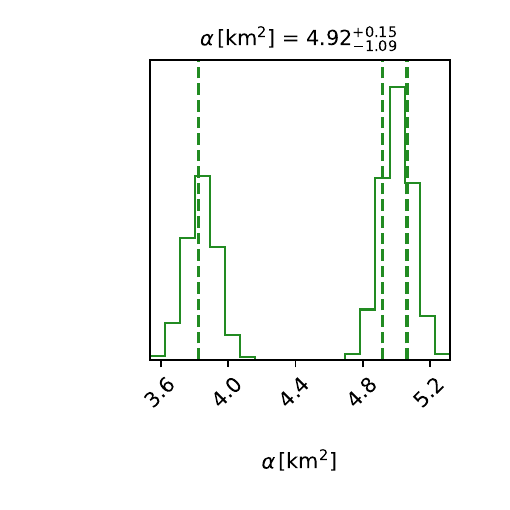}&
\includegraphics[width=0.3\textwidth]{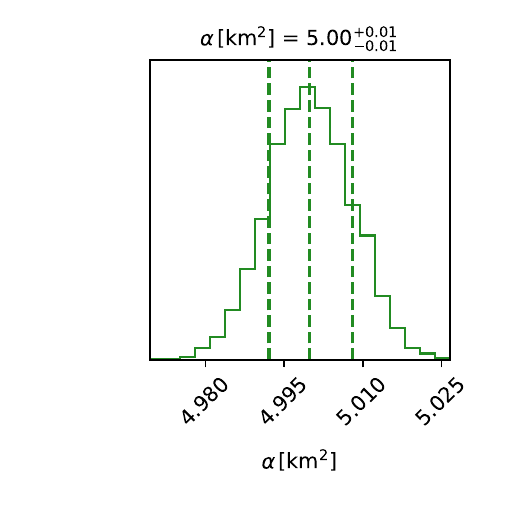}
\includegraphics[width=0.3\textwidth]{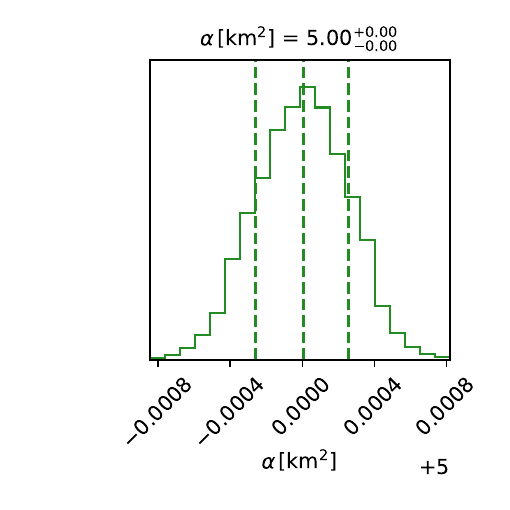} \\
\end{tabular}
\end{center}
\caption{Recovered posteriors of the coupling constant $\alpha$ at A+, Voyager, and 3G sensitivity levels (left, middle, and right columns, respectively), for injections with the SLY4 EoS and $\alpha =[0.05,1,5]$ (top, middle, and last rows, respectively). \bhaskar{For each posterior of $\alpha$, the median and $1 \sigma$ CI are quoted at the top and also shown by the dashed vertical lines. In the case of Voyager and 3G, only the SLY4 EoS is recovered. In the case of A+, the posterior distribution of $\alpha$ also includes a secondary peak corresponding to the SFHO EoS. Nevertheless, for an injection of $\alpha=5\,{\rm km}^2$, the posterior distribution excludes the GR value of $\alpha=0$ at the 3$\sigma$ level.}}
\label{fig:projected_constraint}
\end{figure*}

Based on the projected uncertainty bands on the radius as a function of mass in Figure~\ref{fig:uncertainy_estimation}, we want to determine the anticipated constraints on the EoS uncertainties and the coupling constant $\alpha$. We assume that at a given mass, the uncertainty in radius has a Gaussian distribution, peaked at the injected value ($R$), and the width ($\delta R$) is the maximum of the upper and lower uncertainty values of radius at that given mass ($m$). Therefore, the likelihood has the following form  
\begin{widetext}
\begin{align}
    p(R,\delta R|\mathrm{EoS \, index},\alpha) \propto \int^{1.6M_{\odot}}_{1.0M_{\odot}} dm e^{-\left(\frac{R-R(m,\rm{EoS \, index},\alpha)}{\delta R}\right)^2} \,,
\end{align}
\end{widetext}
and the posterior of the EoS index and $\alpha$ are written as
\begin{widetext}
\begin{align}
    p(\mathrm{EoS \, index},\alpha \mid R,\delta R) \propto p(R,\delta R \mid  \mathrm{EoS \, index},\alpha)p(\mathrm{EoS \, index, \alpha}).
\end{align}
\end{widetext}
Priors on the EoS index and $\alpha$ are kept the same as in our previous investigation. 

In Figure~\ref{fig:projected_constraint}, the posterior distribution of $\alpha$, for the three different chosen injections, is shown for \bhaskar{different} detector sensitivities. 
At A+ design sensitivity, the posterior of $\alpha$ contains two peaks: one around the injected value, whose contribution is coming from the injected EoS, i.e., \bhaskar{SLY4}, and another for a different value of $\alpha$,  whose contribution is coming from SFHO EoS. It is evident that for small injected values of $\alpha=0.05$ and $\alpha=1$, A+ will not have sufficient sensitivity to break the EoS-gravity degeneracy. \bhaskar{However, for larger values, such as $\alpha=5$, we notice that the GR value of $\alpha=0$ is excluded at the $3\sigma$ level, while the EoS SLY4 and SFHO (corresponding to the two peaks in the posterior distribution for $\alpha$)} are very close in the mass-radius diagram (typically within 0.2km of each other for most masses). Hence, if the departure from GR is at or above $\alpha\sim 5$, this could become detectable at A+ sensitivity, within some margin of error.

However, with Voyager and 3G detector sensitivity, we see the posterior of $\alpha$ is peaked around the injected value, and the contribution is solely coming from the injected EoS.

\section{Discussion} 
\label{section: discussion}
In this paper, we delve into the degeneracy between neutron star EoS and gravity within the framework of a 4D Horndeski scalar-tensor theory of gravity, employing a set of 20 realistic EoSs. We establish a Bayesian framework to simultaneously infer the prospective EoS and gravity coupling constant $\alpha$ using both current and forthcoming GW detectors. \bhaskar{We stress that this is a proof-of-principle work since current astrophysical measurements of the tidal deformability using gravitational waves or radii using X-rays have not been interpreted within the framework of the particular theory of gravity we consider. Therefore, it 
It is important to acknowledge the current limitations of our study, which, however, can certainly be alleviated in future work:}

1. The tidal deformability of NSa has not yet been computed within the specific alternative gravity model employed in this paper. This calculation is essential for constraining GW observations more accurately. \bhaskar{In addition, we currently translate the $m-\Lambda$ relation into a $m-R$ relation using universal relations established in GR. So, the constraints presented in Section \ref{subsec:current constraints} should not be taken at face value, but only as indicative results. Nevertheless, the results presented in Section \ref{seq:future} indicate that some first interesting observational constraints could be expected when the GW detectors will be operating at the A+ sensitivity, i.e., several years from now. In this time frame, we hope to alleviate the above limitations and incorporate $\Lambda$ and $R$ measurements consistently within the particular alternative theory of gravity, arriving at quantitative improvements of the constraints presented in Section \ref{section:methods and results}. 

We also stress that the methodology in Section \ref{seq:future} on constraints from future detectors does not incorporate specific tidal deformability measurements. It only assumes that a certain level of accuracy of the measurement of radius as a function of mass (calculated consistently in the framework of a specific theory of gravity) will be achieved as the sensitivity of future detectors will increase with each upgrade or new generation. As detectors will improve, such accuracy levels will be reached eventually (even if on a somewhat different timeline). Other than that, our calculation consistently uses the mass and radius of equilibrium models within the particular theory of gravity we consider.}


2. Presently, we only marginalize over a set of 20 EoSs to deduce the posterior distribution of $\alpha$. However, the constraints we derived here could be influenced by the finite number of EoS candidates considered in this work. In the future, we intend to employ a more comprehensive parameterization of NS EoS that encompasses a broader pressure-density range, allowing us to simultaneously infer the posterior distributions of EoS parameters and $\alpha$.

3. The \bhaskar{projected constraints for future detectors in Section \ref{seq:future}} are illustrated using an expected mass-dependent radius uncertainty at different detector sensitivities. The calculations performed for this purpose are somewhat rudimentary. In reality, a comprehensive analysis should involve injecting multiple binary NS merger events across a range of masses and redshifts to infer both mass and tidal deformability.

Despite the above-mentioned limitations of this study, our findings suggest that with the sensitivity of Voyager or 3G detectors, we may be able to precisely constrain the value of $\alpha$, even if it is very close to GR, and potentially discern the presence of an alternative theory of gravity, thereby breaking the degeneracy between EoS and gravity. At the same time, our study shows that even with the proposed $A+$ upgrade of the 2nd-generation detectors, interesting bounds on deviations from GR may be placed. 

\bhaskar{In future studies, constraints on deviations from GR coming e.g., from BH ringdown (see, e.g. \cite{LIGOScientific:2021sio,Silva:2022srr,2023PhRvD.107l4010E}) should be included in our framework, leading to improved constraints on the EoS of NSs, by breaking the EoS-gravity degeneracy.}

\section*{Acknowledgement} We are grateful to Micaela Oertel for several useful comments and suggestions. We also thank Tim Dietrich, Wolfgang Kastaun and other members of the LVK Extreme Matter working group for comments during a presentation of this work. We are grateful to Andreas Bauswein for providing the EoS tables used in this study and to Christos Charmousis for comments. The research leading to these results has received funding from the
European Union’s Horizon 2020 Programme under the AHEAD2020 project (grant
agreement n. 871158). B.B. acknowledges the support from the Knut and Alice Wallenberg Foundation 
under grant Dnr.~KAW~2019.0112 and the Deutsche 
Forschungsgemeinschaft (DFG, German Research Foundation) under 
Germany's Excellence Strategy – EXC~2121 ``Quantum Universe'' –
390833306. Calculations were performed on the facilities of the North-German Supercomputing Alliance (HLRN) and at the SUNRISE HPC facility supported by the Technical Division at the Department of Physics, Stockholm University.  Virgo is funded through the European
Gravitational Observatory (EGO), by the French Centre National de Recherche Scientifique (CNRS), the Italian Istituto Nazionale di Fisica Nucleare (INFN) and the
Dutch Nikhef, with contributions by institutions from
Belgium, Germany, Greece, Hungary, Ireland, Japan,
Monaco, Poland, Portugal, Spain. 

\bibliography{mybiblio}

\end{document}